\RequirePackage{fix-cm}
\documentclass[smallcondensed]{svjour3}     
\smartqed  
\usepackage[utf8]{inputenc}
\usepackage[english]{babel}
\usepackage{booktabs}
\usepackage{hyperref}
\usepackage[switch,modulo]{lineno}
\usepackage{verbatim}
\usepackage{caption}
\usepackage{subcaption}
\usepackage{graphicx}
\usepackage[shortlabels]{enumitem}
\usepackage{natbib}
\usepackage{tikz}
\usetikzlibrary{positioning}
\usepackage{tikz}
\usepackage[flushleft]{threeparttable}
\usepackage[most]{tcolorbox}
\usepackage{multirow}

\newcounter{chatlinenum}
\begingroup
    \lccode`~=`\[^^M\]
    \lowercase{%
\endgroup
    \def\newchatline#1~{%
        \stepcounter{chatlinenum}%
        \ifodd\thechatlinenum
            \tikz[]{\node[fill=lightgray,chatstyle]{\strut#1\strut};}%
        \else
            \hfill
            \tikz[]{\node[fill=mygreen,chatstyle,align=right]{\strut#1\strut};}%
        \fi
        ~
        \smallskip
    }%
}

\NewDocumentEnvironment{newchat}{}{%
    \setcounter{chatlinenum}{0}
    \begin{minipage}{2.0in}
        \obeylines
        \everypar={\newchatline}
    \end{minipage}
}

\journalname{Empirical Software Engineering}

\newcommand\submittedtext{%
  \footnotesize Preprint. Manuscript accepted at \emph{ 	
Empirical Software Engineering Journal}
  (Springer). Copyright may be transferred without notice, after which this version may no longer be accessible.}

\newcommand\submittednotice{%
\begin{tikzpicture}[remember picture,overlay]
\node[anchor=south,yshift=10pt] at (current page.south) {\fbox{\parbox{\dimexpr0.65\textwidth-\fboxsep-\fboxrule\relax}{\submittedtext}}};
\end{tikzpicture}%
}

\begin{document}

\title{Empirical Evaluation of Taxonomic Trace Links: A Case Study}


\author{Waleed Abdeen       \and
        Michael Unterkalmsteiner \and 
        Peter Löwenadler    \and
        Parisa Yousefi      \and
        Krzysztof Wnuk      
}


\institute{W. Abdeen, M. Unterkalmsteiner, and K. Wnuk \at
              Blekinge Institute of Technology, Sweden \\
              \email{waleed.abdeen@bth.se}           
           \and
           P. Löwenadler, and P. Yousefi \at
              Ericsson, Sweden
}

\date{Received: date / Accepted: date}

\maketitle
\submittednotice

\begin{abstract}
\emph{Context:} Traceability is a key quality attribute of artifacts that are used in knowledge-intensive tasks and supports software engineers in producing higher-quality software. Despite its clear benefits, traceability is often neglected in practice due to challenges such as granularity of traces, lack of a common artifact structure, and unclear responsibility. The Taxonomic Trace Links (TTL) approach connects source and target artifacts through a domain-specific taxonomy, aiming to address these common traceability challenges.
\emph{Objective:} In this study, we empirically evaluate TTL in an industrial setting to identify its strengths and weaknesses for real-world adoption.
\emph{Method:} We conducted a mixed-methods study at Ericsson involving one of its software products. Quantitative and qualitative data were collected across two traceability use cases. We established trace links between 463 business use cases, 64 test cases, and 277 ISO-standard requirements. Additionally, we held three focus group sessions with practitioners.
\emph{Results:} We identified two practically relevant scenarios where traceability is required and evaluated TTL in each. Overall, practitioners found TTL to be a useful solution for one of the scenarios, while less useful for the other. However, developing a domain-specific taxonomy and managing heterogeneous artifact structures were noted as significant challenges. Moreover, the precision of the classifier that is used to create trace links needs to be improved to make the solution practical.
\emph{Conclusion:} TTL is a promising approach that can be adopted in practice and enables traceability use cases. However, TTL is not a replacement for traditional trace links, but rather complements them to enable more traceability use cases and encourage the early creation of trace links.


\end{abstract}

\section{Introduction}\label{sec:intro}

Traceability in software engineering refers to the ability to establish and maintain relationships between artifacts (e.g., requirements, test cases, code) to support tasks such as change impact analysis, compliance verification, and risk assessment~\citep{ieee_glossary_1990,gotel2012traceability}. Traceability is considered a software quality assurance tool~\citep{washizaki_guide_2024}. Traceability is often achieved by establishing and maintaining trace links between development artifacts. These links aid developers in producing correct solutions~\citep{mader_developers_2015}, leading to higher-quality software products~\citep{rempel_preventing_2017}. Moreover, trace links between artifacts increase the value and usefulness of the information they connect. The most common usage scenarios for establishing requirements traceability are finding the origin and rationale of requirements, tracking the implementation state of requirements, analyzing the coverage of requirements in the source code, and developing test cases based on requirements~\citep{bouillon_survey_2013}.  The practical implementation of traceability remains challenging, as reported in recent studies~\citep{fucci_when_2022,maro_tracimo_2022,ruiz_why_2023,mucha_systematic_2024}. Among these challenges, three stand out that we deem to be poorly addressed by traditional trace links that connect source and target artifacts directly.

The first challenge is the difficulty of identifying the right \emph{level of granularity of traces}~\citep{wohlrab_collaborative_2016,maro_tracimo_2022}. Consequently, a trade-off must be made between the usefulness of the links and the effort required to maintain them, especially since artifacts created during software development have varying levels of abstraction~\citep{charalampidou_empirical_2021}. The second challenge is the \emph{lack of a common structure} between artifacts and tools~\citep{fucci_when_2022,mucha_systematic_2024}, which results in large, complex systems with scattered information. When tools lack interoperability and the document structures vary, direct trace links are ineffective. The third challenge is unclear responsibility for establishing traceability~\citep{fucci_when_2022,ruiz_why_2023}. Creating direct links requires knowledge about both source (e.g., requirement) and destination artifacts (e.g., source code), making it often unclear who is responsible for creating and maintaining the links. Furthermore, the creator of the trace link might not be the user of the link, causing traceability to be seen as a burden to the creators. 

Previous work introduced taxonomic trace links (TTL)~\citep{unterkalmsteiner_early_2020}, an indirect traceability mechanism mediated by a domain-specific taxonomy. The central idea of TTL is to exploit the ability of taxonomies to capture shared domain knowledge and use it as an intermediary to connect other artifacts. In our previous work, we conducted a pilot experiment to validate the manual creation of TTL~\citep{unterkalmsteiner_early_2020} and developed a zero-shot classifier for taxonomy-based artifact classification~\citep{abdeen_multi-label_2024,abdeen2025language}. This paper examines the practical utility and deterrents of TTL in industrial settings.

This paper presents the results of an empirical study evaluating TTL’s operational feasibility, strengths, and weaknesses in large-scale software development at Ericsson, a Swedish telecommunications company. Ericsson employs traditional traceability practices, relying on direct artifact links. While these links support basic use cases, the company sought to enhance traceability to enable advanced scenarios that are challenging using traditional trace links. We follow a mixed-methods approach, where we use an exploratory case study as the overarching research method. At the time of conducting this study, there was no well-known taxonomy in the telecommunication domain to classify software requirements and use cases based on identified concepts, which poses a challenge as the taxonomy is an essential part of the TTL approach. Without a taxonomy, we cannot create taxonomic trace links. This study examines the automated creation of a domain-specific taxonomy utilizing Large Language Models (LLMs). The contributions of this paper can be summarized as follows:

\begin{enumerate}
\item Proof-of-concept for TTL deployment in a taxonomy-free domain, including LLM-driven taxonomy generation and automated trace link creation.
\item Quantitative evaluation of TTL’s accuracy in tracing artifacts (requirements and test cases) within a live (deployed) product.
\item Qualitative insights from practitioners on TTL’s utility in two industrial scenarios: software compliance and dependencies identification.
\end{enumerate}

The implications for research are: researchers could benefit from our experience using varied prompts to generate domain-specific taxonomies with generative pre-trained transformers~\citep{radford2018improving}. Further improvements and evaluation of the zero-shot classifier across different datasets are required. Future work could explore automating the capture of domain knowledge in taxonomies and ontologies, which can then be leveraged in TTL to trace and structure development artifacts. The implications for practitioners are that companies aiming to enable traceability scenarios may benefit from creating trace links early using TTL. This study provides a practical guide for implementing TTL-based traceability solutions in real-world software development settings.

The remainder of the paper is structured as follows. Section~\ref{sec:background} contains background information and an explanation of our traceability approach. In Section~\ref{sec:related-work}, we summarize related work. We introduce the research methodology and data collection mechanisms used in this study in Section~\ref{sec:methodology}. Section~\ref{sec:results} contains the results of the study. We discuss the results and their implications in Section~\ref{sec:discussion}. Finally, we conclude the paper and present future work in Section~\ref{sec:conclusion}.

\section{Background}\label{sec:background}

We present background information in this section regarding traceability in software engineering (SE), the proposed approach, taxonomic trace links, and the technical foundation to realize the approach in practice, a zero-shot multi-label classifier.

\subsection{Traceability in SE}

Traceability in SE, according to IEEE, refers to ``the degree to which a relationship can be established between two products of the development process"~\citep{ieee_glossary_1990}. In requirements engineering, traceability refers to ``the ability to describe and follow the lifetime of a requirement in a forward and a backward direction''~\citep{gotel_analysis_1994}. Traceability is often practiced due to its expected benefits for development activities~\citep{bouillon_survey_2013}, as it enables multiple activities, such as change impact analysis and software quality assurance~\citep{gotel2012traceability}. In change impact analysis, traceability between development artifacts helps to understand the relationship between artifacts to identify the impact of a change on the rest of the software~\citep{aung_literature_2020}, e.g., trace links between regulatory requirements and software requirements support identifying the impact of a new regulation on the software requirements~\citep{guo_tackling_2017}. In software quality assurance, trace links between requirements and tests enable practitioners to ensure sufficient test coverage of each requirement~\citep{wohlrab_collaborative_2016}, and tests for risky requirements are prioritized.

The process of trace link creation has been categorized into two basic approaches~\citep{gotel2012traceability}. In trace capture, links are created concurrently with the artifacts that are associated with each other~\citep{ramesh_toward_2001}. In trace recovery, existing artifacts are analyzed to identify associations between them~\citep{cleland-huang_utilizing_2005}. Trace capture has the advantage that it is easier to validate trace link fidelity while the artifacts are created, involving the creators of the artifacts, as opposed to recovery, where trace links are typically not recovered by the artifact creators~\citep{wohlrab_collaborative_2016}. Trace recovery has the advantage that trace links can be created on demand and does not need any upfront investment in creating links that may not be used~\citep{cleland-huang_utilizing_2005}.

Early work by Kaindl~\citep{kaindl_missing_1993} proposed using taxonomies derived from software design to classify requirements and establish hierarchical relationships between entities. While this approach aimed to improve the organization and navigation of requirements, it focused on structuring domain objects rather than creating trace links between artifacts. In contrast, our work leverages domain-specific taxonomies to derive Taxonomic Trace Links (TTL)~\citep{unterkalmsteiner_early_2020}.

\subsection{Taxonomic Trace Links}

\begin{figure}
\centering
\begin{subfigure}{.4\textwidth}
  \includegraphics[width=1\linewidth]{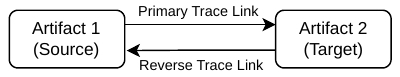}
  \caption{Traditional trace link (adapted from~\citep{gotel2012traceability})}
  \label{fig:traditional}
\end{subfigure}
\begin{subfigure}{.45\textwidth}
  \includegraphics[width=1\linewidth]{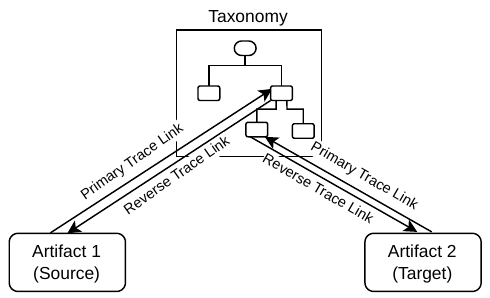}
  \caption{Taxonomic trace link}
  \label{fig:taxonomic}
\end{subfigure}
\caption{Traditional vs. taxonomic trace links}
\label{fig:tvst}
\end{figure}

A trace is formally defined as a triplet comprising a source artifact, a target artifact, and a bidirectional link connecting them~\citep{gotel2012traceability}. Traditional direct trace links explicitly connect artifacts, as illustrated in Figure~\ref{fig:traditional}. In contrast, Taxonomic Trace Links (TTL)~\citep{unterkalmsteiner_early_2020} introduce an indirect connection mediated by a domain-specific taxonomy, a set of concepts specific to the domain arranged in a hierarchy. An example of such a taxonomy is the Banking Industry Architecture Network (BIAN) architectural reference model (service landscape)~\footnote{https://bian.org/servicelandscape-12-0-0/}, which defines a service-oriented architecture (SOA) for the banking industry and serves as a reference to build a banking system. As shown in Figure~\ref{fig:taxonomic}, TTL associates source and target artifacts using classes from a taxonomy, enabling traceability on various levels of abstraction.

A taxonomy in this context is a hierarchical structure of nodes, each representing a domain concept. Nodes minimally include a title and may include descriptions or synonyms. Each node (except the root) has one parent and zero or more children. The taxonomy captures one or more dimensions of the problem domain. For example, consider we need to create a trace link between the requirement \emph{R1- The system shall allow a subscriber to initiate a voice call to another subscriber by dialing their phone number}, and the test cases \emph{TC1 – Verify successful call setup between two available subscribers} and \emph{TC2 – Verify call attempt fails when called subscriber is unavailable}. Using a taxonomy containing, among others, the classes: \emph{A1 - voice call} and \emph{B1 - subscriber}, we can say that R1, TC1, and TC2 can be classified using the classes \emph{A1} and \emph{B1}. Consequently, the following trace links pairs [R1 <-> TC1], R1 <-> TC2] exist between the artifacts, as they are both classified using the same classes.

We argue that TTL addresses three key challenges of traditional traceability:
\begin{enumerate}
    \item Difficulty identifying the right granularity of traces: Development artifacts are typically created at different levels of abstraction, requiring decisions about trace link granularity~\citep{wohlrab_collaborative_2016,maro_tracimo_2022}. TTL addresses this by creating trace links to a common domain taxonomy with multiple abstraction levels, allowing trace link usage at different granularities through the taxonomy's hierarchical structure.    
    \item Lack of common structure between artifacts: Software development involves collaboration between multiple teams working on different activities, often resulting in artifacts and tools that lack a unified structure unless explicitly enforced~\citep{fucci_when_2022,mucha_systematic_2024}. TTL introduces tracing through a domain-specific taxonomy that can serve as a common model to structure both tools and artifacts.
    \item Unclear responsibility for traceability~\citep{fucci_when_2022,ruiz_why_2023}: Traced artifacts (e.g., requirements, source code, and test cases) are produced at different stages of development. Creating direct trace links requires both source and target artifacts to exist, often delaying trace link creation until later stages~\citep{fucci_when_2022,ruiz_why_2023}. This results in unclear responsibility for creating and maintaining links. TTL enables each stakeholder to take ownership of traceability for their artifacts by creating links to the taxonomy, resulting in trace links to all other traced artifact types in the development model.
\end{enumerate}

Implementing \emph{taxonomic-trace links} in practice requires support to create and maintain these links. Manually classifying artifacts with large taxonomies is challenging and error-prone~\citep{unterkalmsteiner_early_2020}. Thus, a classifier is required to support practitioners by recommending the top classes from the taxonomy for the traced artifacts to implement TTL. 

\subsection{Zero-Shot Classifier}

Using automated approaches to implement \emph{taxonomic-trace links} is particularly difficult when labeled data are scarce, as is often the case for RE tasks~\citep{ferrari2017natural}. Supervised machine learning (ML) approaches~\citep{kurtanovic_automatically_2017,hey_norbert_2020} require a sufficient amount of training data for each class. Domain-specific taxonomies may have hundreds or thousands of classes, making it close to impossible to create a balanced and sufficiently large dataset for training.

Zero-shot learning (ZSL) for classification is the method of learning a classification without training data being available for all classes~\citep{larochelle_zero-data_2008}. In the context of NLP, ZSL leverages the usage of pre-trained models to predict unseen classes~\citep{xian_zero-shot_2017,wang_survey_2019}. ZSL does not require a labeled dataset for training, and transferring a classifier to a new domain does not require retraining~\citep{rezaei_zero-shot_2020}. However, to evaluate the performance of a zero-shot learner, labeled data is still required, which can be significantly smaller than the training data to train a supervised machine learning model.

In previous studies~\citep{abdeen_multi-label_2024,abdeen2025language}, we introduced and evaluated a zero-shot requirements classifier that assigns classes from a domain-specific taxonomy to natural language requirements. This classifier semi-automates the classification of artifact elements (e.g., requirements and test cases) using the taxonomy, thereby reducing the effort required to create Taxonomic Trace Links (TTL). A ZSL classifier is unlikely to reach 100\% precision, i.e. the result contains false positives. Hence, a human in the loop is required as the final judge to vet the correctness of the links.

\begin{figure}
    \centering
    \includegraphics[width=\linewidth]{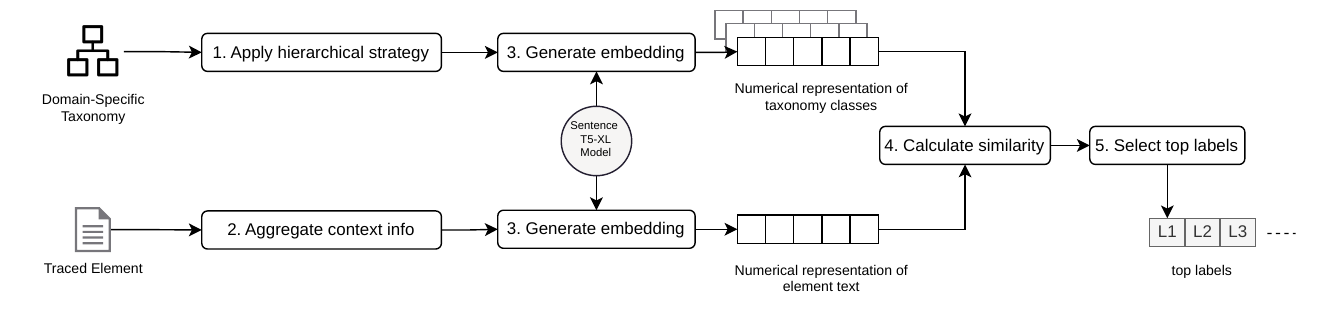}
    \caption{Zero-Shot Classifier~\citep{abdeen2025language}}
    \label{fig:zsc}
\end{figure}

Figure~\ref{fig:zsc} illustrates the architecture of the zero-shot classifier. The classifier uses natural language processing and exploits the semantic knowledge captured in language models to, in essence, identify similarities between an artifact and a domain concept. The classifier takes as input a domain-specific taxonomy and some text that we want to classify. The text can originate from, for example, a requirement, a source code comment or documentation, a bug report, or a test case. The classifier pre-processes both inputs and then feeds the processed text to a sentence T5-Xl transformer, a language model developed by Google and fine-tuned for translation tasks~\citep{ni_sentence-t5_2021}. The T5 model family is referred to as sequence-to-sequence model and contains both an encoder and a decoder. Although these models are trained on different tasks, they can generate sentence embeddings using the encoder that are useful to find semantic similarity~\citep{abdeen2025language}. The language model generates embeddings, a numerical representation, for each class in the taxonomy and the text to classify. Finally, cosine similarity is calculated between the element text embedding and the embeddings of each of the classes. The output of the classifier is a list of labels ordered by the similarity score between the input text embeddings. For a detailed explanation of the architecture and implementation, refer to~\citep{abdeen2025language}. 

In our previous study~\citep{abdeen2025language}, the classifier achieved a recall of 78\%, indicating strong coverage of relevant classes, and a precision of 7\%, reflecting a high rate of false positives. While the high recall minimizes missed classifications, the low precision necessitates human validation to filter out incorrect class assignments before finalizing trace links. This raises the question of whether the performance of the zero-shot classifier is adequate for practical applications.

\section{Related Work}\label{sec:related-work}

Traceability has been extensively studied in software engineering research~\citep{cleland-huang_utilizing_2005,di_improving_2009,mahmoud_semantic_2012,panichella_when_2013,guo_semantically_2017,schlutter_trace_2020,zhang_ealink_2023,keim_recovering_2024}. Most studies have focused on trace link recovery using information retrieval (IR) techniques, where links are established during later development stages or when specifically needed. The primary motivation for IR approaches has been to improve efficiency and reduce the effort required for trace link creation and maintenance.

Despite their promise, IR techniques face limitations when used for traceability, due to semantic gaps between traced artifacts. Researchers have attempted to address these shortcomings through various methods, some examples are: Bayesian classification~\citep{di_improving_2009}, semantic relatedness measures~\citep{mahmoud_semantic_2012}, user feedback integration~\citep{panichella_when_2013}, query augmentation~\citep{guo_tackling_2017}, semantic relationship graphs~\citep{schlutter_trace_2020}, and transitive link analysis~\citep{keim_recovering_2024}. These improvements have demonstrated effectiveness, with recent approaches achieving F1-scores between 82\%--98\%~\citep{keim_recovering_2024}. These studies focus mainly on the technical solution to improve the performance of direct trace link creation using IR. However, our study mainly focuses on evaluating TTL practicality in realistic settings.

Wang et al.~\citep{wang_tagging_2015} proposed the use of assisted tagging during tracing and developed a prototype as an eclipse plugin to evaluate the effectiveness of the approach and the benefits of tags. The main idea is to allow practitioners to tag text and use cases using keywords. Twenty-eight engineering students participated in the evaluation of a software from the medical domain. The results suggest that tagging can be adopted by humans in tracing and improves the accuracy of the final traces. However, the absence of practitioners in the study limits the generalizability of the findings to real-world industrial settings.

Klimpke and Hildenbrand~\citep{klimpke_towards_2009} conducted five case studies on companies from different sectors and of different sizes (200-30000 employees). The focus of the study was to assess the current traceability practices and identify requirements for end-to-end traceability. Based on their results, traceability is challenging to adopt despite the existence of tools that support it, mainly due to the heterogeneity of existing development tools, which makes integration difficult. Furthermore, traceability tools adoption could be hindered due to the lack of support for all development phases. 

Maro et al.~\citep{maro_tracimo_2022} introduced TracIMo, a methodology for incrementally deploying traceability in a financial domain company. Their iterative approach emphasized tailoring traceability to organizational needs but revealed challenges such as defining trace granularity and managing additional artifacts used to support tracing. TTL introduces indirection in traceability through shared taxonomy classes, potentially addressing the granularity of traces challenge.

Recently, researchers have investigated using LLMs to recover trace links between various software artifacts. Lin et al.~\citep{lin_traceability_2021} have proposed an approach to recover trace links between artifacts by fine-tuning BERT-based models using a labeled dataset. They evaluate their approach using experiments to recover trace links between requirements and source code in open source projects, and show significant improvement over an information retrieval approach. Rodriguez et al.~\citep{rodriguez_prompts_2023} have studied using LLMs to generate traceability links. In their study, they focus on prompt engineering and illustrate how small changes in the prompts could significantly affect the output of the language model. Our approach differs from those as we advocate for a zero-shot learning approach without any fine-tuning or the need for any training data.

Although software engineering researchers have proposed solutions to address requirements traceability challenges, improved on the technical solution of trace links creation, and evaluated the solution in controlled environments, the empirical evaluation of traceability approaches in a realistic setting is still limited~\citep{klimpke_towards_2009,wang_tagging_2015,maro_tracimo_2022}.

Prior to this study, we evaluated Taxonomic Trace Links (TTL) through controlled experiments focused on: manual trace link creation~\citep{unterkalmsteiner_early_2020}, investigations of the classification system's structural properties~\citep{abdeen_multi-label_2024}, and validation of zero-shot classifiers for artifact classification~\citep{abdeen2025language}. However, these evaluations lacked testing in realistic industrial settings. Our current work addresses this gap by investigating the operational feasibility of TTL for large-scale, regulated telecommunications software systems that require traceability.

\section{Research Methodology}\label{sec:methodology}

In this study, we empirically evaluate the feasibility, performance, and practical utility of  Taxonomic Trace Links (TTL) for tracing requirements to other artifacts in the context of software development within industrial settings. The study was conducted at Ericsson, a telecommunications company in Sweden, focusing on one of its software development units. The study has three primary objectives:

\begin{enumerate}[O1]
\item To evaluate the feasibility of the TTL approach in a context without pre-established taxonomy.
\item To evaluate the effectiveness of TTL in tracing software requirements in practice.
\item To evaluate the strengths and weaknesses of TTL for different use cases of requirements traceability.
\end{enumerate}

To address these objectives, we formulate the following research questions:

\begin{enumerate}[RQ1]
    \item \emph{How feasible is it to implement TTL in a domain with no well-known established taxonomy?}
    \item \emph{What is the performance of the ZSL classifier in recovering trace links between use cases and test cases, measured in terms of precision, recall, and F1-score?}
    \item \emph{What is the practical utility of TTL for supporting different traceability use cases?}
\end{enumerate}

The primary research method employed in this study is a case study, guided by established case study research guidelines~\citep{runeson_guidelines_2008}. We adopted an iterative approach for the design and execution of the case study, illustrated in Figure~\ref{fig:method}. The process began with defining the study's goal, objectives, and primary research questions (RQ1, RQ2 and RQ3). Then, after the initial assessment of the current traceability practices, we defined subquestion RQ2.1 .

\subsection{Case Description}
We conducted a case study at Ericsson, focusing on charging management --- the systems responsible for measuring customer network usage, applying tariff rules, generating invoices, and ensuring regulatory compliance for payment processing. This domain handles complex requirements involving real-time transaction processing and integration between network elements and business support systems. The unit of analysis was a customer-facing product. We specifically examined the requirements unit, which oversees end-to-end requirement development from initial business needs through to mature use case specification. Additionally, we involved the testing unit responsible for writing and running test cases for each of the written use cases. Two champions from the requirement unit were assigned to this study, who eased access to internal company data and helped us understand the processes and context. One of the champions participated in the focus group sessions. Our motivation for selecting the company and product is twofold:

\begin{enumerate}
    \item Regulated Domain: Telecommunications products must comply with specific international standards, such as ISO 14452:2012~\citep{ISO_144522012}, and 3GPP 32.240 \citep{3GPP32.240}. These standards must be explicitly connected to the internal development documentation to ensure compliance and facilitate audits.
    \item Large-Scale Product: The product's size and complexity, with numerous features and components, make it challenging to maintain high software quality. A robust and diverse set of trace links is essential for effective bug tracing and change impact analysis, which are critical for managing such a large system.
\end{enumerate}

Traditional direct trace links have already been implemented in the product, where source and target artifacts are connected using unique identifiers. However, these links were primarily created at a high level of abstraction and could benefit from greater granularity to enhance their usefulness for engineers and enable more detailed tracking of development progress. Ericsson continuously seeks to improve its processes through research and development initiatives, making it an ideal context for evaluating innovative traceability approaches.

\subsection{Case Study Design}

\begin{figure}
    \centering
    \includegraphics[width=\linewidth]{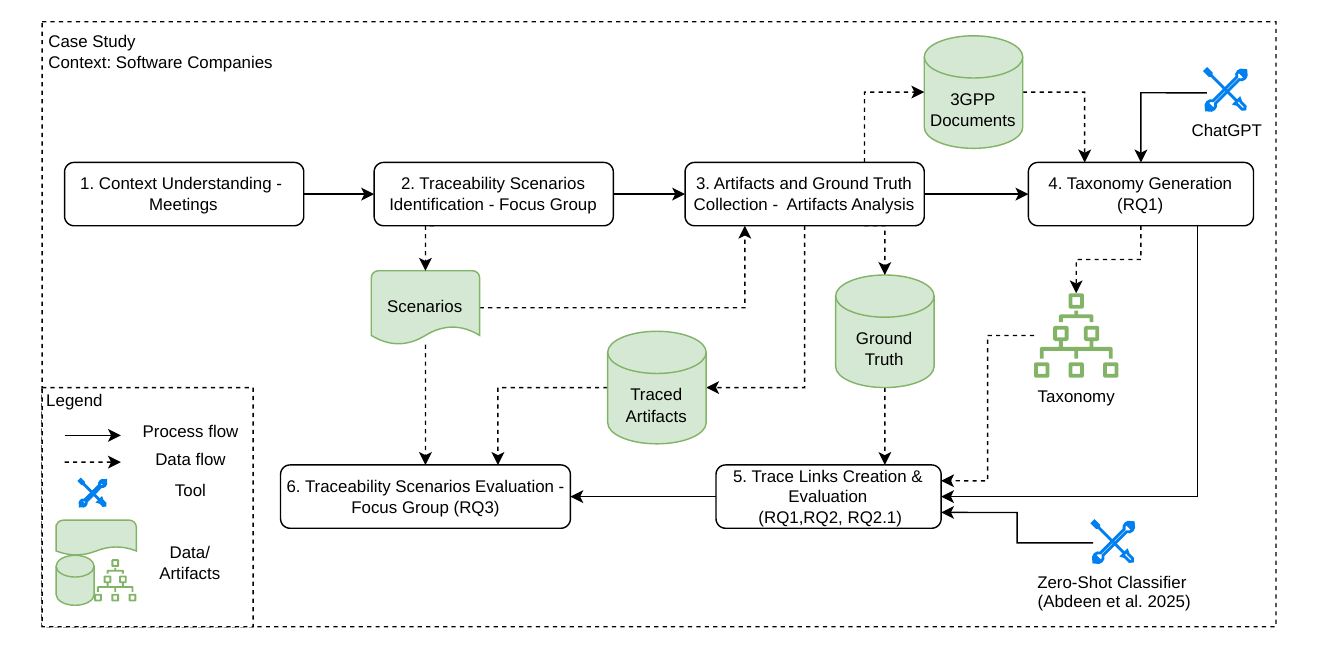}
    \caption{Case Study Design}
    \label{fig:method}
\end{figure}

Figure~\ref{fig:method} depicts the steps we followed to conduct this study and address the research questions. We describe each of the six steps next.

\subsubsection{Context Understanding}

We began by conducting a series of context understanding meetings with the champions at the company. The goal of these meetings was to understand the context, the development process and current traceability practices. During this step, we gained access to systems, artifacts, and further relevant stakeholders at the company. 

\subsubsection{Traceability Scenarios Identification}\label{sec:scenarios-focus}

We continued by identifying traceability scenarios that stakeholders perceived as valuable and are currently not well supported with traditional trace links. To achieve this, we designed and conducted focus group sessions following established guidelines~\citep{shull_guide_2008}. We invited six participants with diverse roles, as detailed in Table~\ref{tab:focus_one_participants}, to the session. Prior to the session, participants were asked to complete a questionnaire to assess the current state of traceability at Ericsson.

Through this step, we identified multiple traceability scenarios and collaborated with stakeholders to prioritize two key scenarios based on their relevance and impact on the development process. The prioritization was finalized in meetings with two primary stakeholders, who evaluated the scenarios based on feasibility, alignment with research objectives, and practical relevance. These scenarios informed the formulation of sub-research question RQ2.1.

\begin{table}[tbh]
    \centering
    \caption{Participants in Scenarios Identification Focus Group}
    \begin{tabular}{ccc}
        \toprule
        Participant & Role & Experience (years) \\
        \midrule
        1 & Chief System Architect & 15 \\
        2 & Solution Architect & 5\\
        3 & Line Manager (Solution Architect) & 10 \\
        4 & Technical Manager (Test) & 12 \\
        5 & Release Architect & 5 \\
        6 & Solution Architecture & 20+ \\
        \bottomrule
    \end{tabular}
    \label{tab:focus_one_participants}
\end{table}

\subsubsection{Artifacts and Ground Truth Collection}\label{sec:artifacts-collection}

We collected and analyzed artifacts at the company to support the subsequent steps. During this step, we collected three main artifacts. 1) 3GPP documents~\footnote{https://www.3gpp.org/}: a set of standards that software products in the telecommunication domain should adhere to. These standards were used to build the taxonomy. A complete list of these documents is available in the Appendix~\ref{sec:appendix-3gpp}. 2) Ground truth: a set of 64 trace links connecting the requirements with the test cases, which were created manually by the testers when they wrote the test cases. These links are added as annotations to the test method using the requirement document ID. Each test case is connected to one requirement document only, out of hundreds of documents. The ground truth was used to measure the performance of the zero-shot classifier and select the number of labels per artifact (K), the number of matching labels to consider a trace link exists and select the language model used in the classier that leads to the best performance results, 3) Traced artifacts: artifacts collected based on the identified scenarios and containing requirements, tests, and standards. All these collected artifacts were analyzed and then used in subsequent steps of this study.

\subsubsection{Taxonomy Generation}
Ericsson did not use a domain-specific taxonomy to classify requirements documents prior to this study. Instead, the requirements engineers cluster the specifications (on a high level) based on the product aspect that the specification focuses on, primarily to organize requirements and the software architecture. Therefore, we needed to develop a taxonomy to classify the traced artifacts. We initially searched the literature, using Google Scholar and gray literature using Google, for publicly available taxonomies to classify software requirements and use cases in the telecommunication charging domain, but were unable to find one that suited our purpose. For example, the Network and Services Management Taxonomy by IEEE~\citep{dos2016taxonomy} covers a broader area (multiple aspects) from the telecommunication domain; however, the taxonomy contains only two levels --- when used for creating trace links, this results in abstract trace links. Another example is the Information Framework (SID) by TMForums~\citep{tmforumInformationFramework}, which is a framework designed for the telecommunications and digital service provider domain. The frameworks cover multiple sub-domains (e.g., customer, product, and resources); however, they are presented on a high level. Consequently, we developed a taxonomy for the purpose of this study.  

We evaluated two automated taxonomy generation approaches to build a taxonomy for the telecommunication domain: TaxoGen~\citep{zhang_taxogen_2018}, an unsupervised method that constructs topic taxonomies through embedding and clustering of domain corpora, and TaxoCom~\citep{lee_taxocom_2022}, a taxonomy completion approach that expands from seed terms to better align with stakeholder needs. However, both methods produced limited taxonomies for the charging and billing domain, yielding a small number of classes (up to 52) with only a few being relevant. By examining these classes, we found that they are included mainly due to their frequent use in the input documents (e.g., 5G and system). The remaining classes were not specific to the domain (e.g., accessible, unit, and information). Furthermore, the parent-child relationship between the nodes was inaccurate.

Recent advances in Generative AI (GenAI) have demonstrated effectiveness across software engineering tasks, including requirements analysis~\citep{fantechi_inconsistency_2023}, coding~\citep{rajbhoj_accelerating_2024}, and testing~\citep{aleti_software_2023}. Given these successes, we employed ChatGPT (version 4o), one of the top-performing GenAI models at the time of our study, for domain-specific taxonomy generation. This choice was motivated by its state-of-the-art performance across diverse NLP tasks and its ability to handle complex domain-specific queries. Details about the developed approach are presented in Section~\ref{sec:taxonomy-results}.

\subsubsection{Trace Links Creation \& Evaluation}
To create taxonomic trace links between software development artifacts, it is necessary to classify both the source and target artifacts using the same taxonomy~\citep{unterkalmsteiner_early_2020}. For this purpose, we employed a zero-shot classifier based on language models, which we introduced in a previous study~\citep{abdeen2025language}. The classifier analyzes the taxonomy and input text of the artifact to be classified and recommends top-k labels from the taxonomy as a classification for the input text. K is an adjustable configuration parameter that allows for flexibility in implementing the classifier in different contexts, where the taxonomy size and classification results may vary. Selecting K requires running the classifier on a test dataset and then adjusting it to get the best possible performance.

\begin{figure}
    \centering
    \includegraphics[width=1\textwidth]{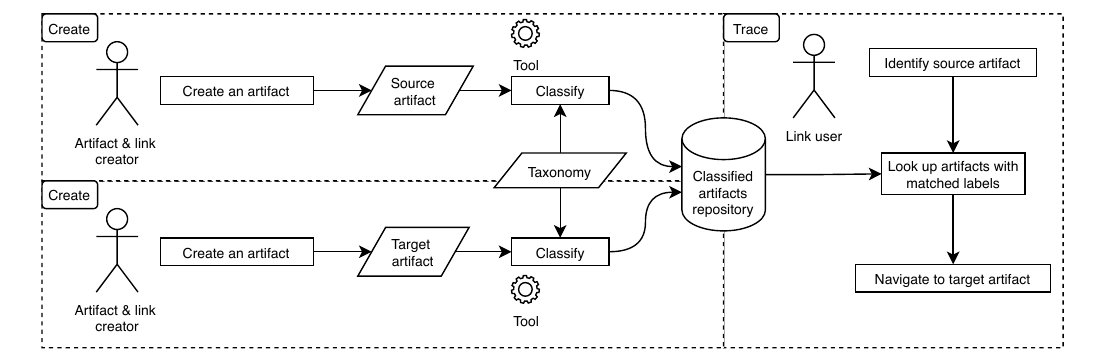}
    \caption{TTL creation and usage}
    \label{fig:ttl_create_and_trace}
\end{figure}

Furthermore, we evaluated the performance of the classifier on a ground truth, which we collected during artifact analysis (Section~\ref{sec:artifacts-collection}). We used precision, recall, and F1-score metrics to measure the classifier's performance. Figure~\ref{fig:ttl_create_and_trace} depicts the process of creating TTLs.

\subsubsection{Traceability Scenarios Evaluation}\label{sec:evaluation-focus}
To evaluate the perceived practical utility of these trace links in various scenarios, we designed and conducted a set of focus group sessions. These sessions focused on evaluating the strengths and weaknesses of TTL in different scenarios.
    
We conducted two focus group sessions, each involving two or three participants with relevant experience to the specific scenario. Three of the participants, who work with requirements, were involved in the focus group session where we identified the traceability scenarios (Section~\ref{sec:scenarios-focus}), while the other participants were from software compliance unit. The sessions were facilitated by the first author. Before each session, we generated the trace links using the classifier to classify the elements of the traced artifacts according to the taxonomy presented in Section~\ref{sec:taxonomy-results}. The trace links were then deduced based on partial matching of the recommended labels of each artifact. A partial matching exists when at least one of the recommended labels is associated with both artifacts. We classified and traced artifacts that we collected during artifact collection and analysis (Section~\ref{sec:artifacts-collection}) based on the selected scenarios. 

Each focus group session lasted two hours. We started by introducing background information and task descriptions. Then, we asked the participants to perform the task on three sets of trace links, allocating 30 minutes to each set. In the end, we asked the participants to complete a brief questionnaire to gather their feedback on the solution. The questionnaire used in the focus group session is provided in Appendix~\ref{sec:appendinx-eval}.
    
\begin{table}[tbh]
    \centering
    \footnotesize
    \caption{Participants in Traceability Evaluation Focus Group}
    \begin{threeparttable}
    \begin{tabular}{ccp{0.32\textwidth}c}
        \toprule
        Participant & Scenario & Role & Experience\tnote{a}  \\
        \midrule
        1 & Software Compliance & Release Operating System Manager & 20+ \\
        2 & Software Compliance & Senior Security Architect & 14 \\
        3 & Dependency Identification & Chief System Architect & 15 \\
        4 & Dependency Identification & Release Architect & 5 \\
        5 & Dependency Identification & Solution Architect & 20+ \\
        \bottomrule
    \end{tabular}
    \begin{tablenotes}
        \item [a] In years
    \end{tablenotes}
    \end{threeparttable}
    \label{tab:focus_two_participants}
\end{table}

\subsection{Data Analysis}
We used thematic coding to analyze the qualitative data from both focus group sessions. From the first focus group session (traceability scenarios identification), the first author coded the session notes taken by the facilitator, the brainstorming notes provided by the participants, and the transcription of the session. The coding primarily consisted of the code ``challenge'' and a code for each identified challenge. These codes were then discussed between the authors until consensus was reached, and then presented to the company champions to verify and prioritize the challenges in feedback sessions. From the second focus group session (traceability scenarios evaluation), we coded the questionnaire answers provided by the participants at the end of the session. The codes contained two main codes: \emph{weakness} and \emph{strength}, and specific instances of each. We grouped together similar codes into themes. We used these themes to analyze and report the results (Section~\ref{sec:results}). For example, during the first focus group session (traceability scenarios identification), different people used different formulations to refer to the same scenarios. Consequently, using these themes allowed us to report the combined views of the participants.

\section{Results}\label{sec:results}

We present the results of both focus group sessions, the taxonomy generation, and trace links creation and evaluation. 

\subsection{Traceability Scenarios Identification}\label{sec:identified-scenarios}
We identified five potential scenarios and prioritized two key scenarios for traceability at Ericsson, as described below:

\begin{enumerate}[A]
    \item \emph{Software Compliance}: Ericsson's software product must comply with so-called ``General Product Requirements'' (GPRs) that contain domain-specific standards as well as quality aspects that need to be fulfilled by products with relevant capabilities. When specifying a new requirement based on customer needs, compliance engineers often rely on their product knowledge to identify relevant GPRs, which can be time-intensive and error-prone. At the time of conducting the study, 277 GPRs existed that the compliance engineers were required to check. A GPR has a title, text (a couple of sentences to a paragraph), a rationale, and a classification based on 17 categories. In this scenario, we aim to assist compliance engineers in identifying applicable GPRs for specific business use cases (BUCs), thereby streamlining the compliance verification process.

    \item \emph{Dependencies Identification}: Product owners and requirements engineers must identify dependencies and potential conflicts between BUCs when authoring a new BUC. Due to the large-scale nature of the software and the high volume of documented BUCs, manually identifying relevant BuCs is time-intensive, especially when multiple stakeholders are involved in documenting or updating a BUC. Over a two-year period, 462 BUCs were written by the requirements unit. Each BUC can consist of descriptions ranging from 1 to 3 pages. Identifying dependency relationships between BUCs without any aid requires, therefore, considerable time. In this scenario, we aim to support product owners in automatically correlating related BUCs to flag dependencies and conflicts.  

\end{enumerate}

The listed scenarios are those that were prioritized by stakeholders and deemed useful for the requirements unit. 

\subsection{Taxonomy Generation}\label{sec:taxonomy-results}

To generate a domain-specific taxonomy, we used ChatGPT 4-o. We began by prompting ChatGPT to develop a taxonomy for the telecommunication charging domain, without any further context. The initial prompts yielded a high-level summary of what such a taxonomy might include; however, this result was insufficient for our use case, and further improvement on the prompts was required. To refine the taxonomy, we iteratively improved the prompts based on the model's responses and its alignment with our expectations (a set of concepts from the telecommunication charging domain, arranged in a hierarchy, each has a unique Id each); moreover, we followed the lessons learned from \citet{rodriguez_prompts_2023}, mainly that, small modification can lead to significant differences in the output, and being more specific gives better results.

\subsubsection{Prompt engineering} We developed and used three strategies to generate a domain-specific taxonomy.

\begin{enumerate}
    \item \emph{All-at-Once Strategy}: We instructed the LLM to generate the taxonomy in a single step, specifying the domain, nodes to include, and desired granularity. The LLM was prompted to first request the user’s preferred granularity level before generating the full taxonomy. Figure~\ref{fig:strategy1} shows the detailed instructions.
    
    \begin{figure}[h!]
        \centering
        \begin{tikzpicture}[node distance=0.5cm and 0.5cm]
            \node[draw, fill=green!10, rounded corners, anchor=south east, text width=0.6\textwidth, align=left] (msg1) at (0,0)
            {You are an expert in the telecommunication domain. Your task is to build a taxonomy specific to the telecommunication charging management domain. Each node in the taxonomy should represent an entity from the domain. Start by asking the user the level of granularity needed. Then build the taxonomy with the required level of granularity. Make sure to give a unique numerical ID for each node};
            \node[above=0.2cm of msg1.north west, anchor=south west] {\textit{Instructions}};
        \end{tikzpicture}        
        \caption{Instructions provided to ChatGPT: All at once strategy}
        \label{fig:strategy1}
    \end{figure}

    This is the simplest approach, where one asks the model as specifically as possible what they want and how the output should look. We observed that the generated taxonomy only contains eight top-level classes, and each node has a maximum 2-3 children. Even though we repeated the prompt multiple times, the model stopped generating text after a fixed number of tokens/lines of text.
    
    \item \emph{Bottom-Up Strategy}: Building on the first strategy, we instructed the LLM to iteratively construct the taxonomy from leaf nodes upward. The model first generated leaf-level entities, then prompted the user to abstract these into higher-level nodes until reaching the root. Figure~\ref{fig:strategy2} details the instructions.      
    \begin{figure}[h!]
        \centering
        \begin{tikzpicture}[node distance=1cm and 1cm]
            \node[draw, fill=green!10, rounded corners, anchor=south east, text width=0.6\textwidth, align=left] (msg1) at (0,0)
            {You are an expert in the telecommunication domain. Your task is to build a taxonomy specific to the telecommunication charging management domain. Each node in the taxonomy should represent an entity from the domain. You need to use a bottom-up approach. First, provide a list of the bottom-level nodes, then ask the user if they want to abstract from these nodes further until they say stop or you reach the root node of the taxonomy. Make sure to give a unique numerical ID for each node};
            \node[above=0.2cm of msg1.north west, anchor=south west] {\textit{Instructions}};
        \end{tikzpicture}        
        \caption{Instructions provided to ChatGPT: Bottom-up strategy}
        \label{fig:strategy2}
    \end{figure}

    Using this strategy allowed us to generate more nodes, especially after continuously asking the model, \emph{are there more nodes?} However, the model failed to abstract from the child nodes correctly, and it could forget to include some of the nodes identified in the first step in the output.
    
    \item \emph{Level-Branch Strategy}: In this approach, the LLM generated the taxonomy hierarchically, starting with top-level nodes and progressively decomposing them into sub-nodes based on user-specified granularity. Figure~\ref{fig:strategy3} provides the instructions to the LLM.
    
    \begin{figure}[h!]
        \centering
        \begin{tikzpicture}[node distance=1cm and 1cm]
            \node[draw, fill=green!10, rounded corners, anchor=south east, text width=0.6\textwidth, align=left] (msg1) at (0,0)
            {You are an expert in the telecommunication domain. Your task is to build a taxonomy specific to the telecommunication charging management domain. Each node in the taxonomy should represent an entity from the domain. Start by identifying the top-level nodes, then ask the user if they want to break down a specific node and the required depth level (e.g., 2,3,4, etc.), while considering the top level as depth level 1. Make sure to give a unique numerical ID for each node};
            \node[above=0.2cm of msg1.north west, anchor=south west] {\textit{Instructions}};
        \end{tikzpicture}        
        \caption{Instructions provided to ChatGPT: Level-Branch strategy}
        \label{fig:strategy3}
    \end{figure}

    This strategy produced taxonomies with better structure and a higher number of classes than those generated by previous strategies. It overcame the context window limit of the model (output limit) from the first strategy, and avoided misunderstanding of how abstraction is performed and parent nodes are created from the second strategy. However, many of the generated nodes were redundant in different branches.
\end{enumerate}

These strategies were the result of iterative improvement on the prompt based on the model output. We stopped at the last strategy as we had reached a stage where the generated taxonomy appeared to be useful for our work, as perceived by the Chief System Architect of the product, who reviewed the output.

\begin{table}[tbh]
    \centering
    \caption{Generated Taxonomies for the Telecommunication Charging Domain with corresponding node count (n), leaf nodes (l), categories (c) and depth level (d)}
    \begin{threeparttable}
    \begin{tabular}{lccc}
        \toprule
         Input Data $\setminus$ Strategy & All at Once &  Bottom-Up  & Level-Branch \\
         \midrule
         \multirow{2}{*}{no} & $T_{1a}$: n=68, l=39, & $T_{1b}$: n=75, l=48, & $T_{1c}$: n=782, l=528,\\
         & c=29, d=4 & c=27, d=4  &  c=254, d=4\\
         \multirow{2}{*}{29x 3GPP standards docs}  & $T_{2a}$: n=91, l=62, & $T_{2b}$: n=62, l=51,& $T_{2c}$: n=859, l=581, \\
         & c=29, d=4 & c=11, d=3 & c=278, d=4 \\
         \multirow{2}{*}{32.260 document (3GPP)} & $T_{3a}$: n=71, l=49, & $T_{3b}$: n=70, l=48, & $T_{3c}$: n=876, l=593, \\
         & c=22, d=4 & c=22, d=3 & c=283, d=4\\
         \bottomrule
    \end{tabular}
    \end{threeparttable}
    \label{tab:taxonomies}
\end{table}

\subsubsection{Data Sources}
We experimented with three data sources to provide context for the GenAI model when generating the taxonomy. Without providing any additional data, the model generated a taxonomy that is somehow relevant to the telecommunication billing and charging domain. However, it was not particularly useful for classifying the software artifacts of the product used for investigation. Therefore, we opted to provide data sources that are relevant to the product. These data sources were recommended by the Chief System Architect of the product, based on their experience in regulations and standards.

\begin{enumerate}
    \item \emph{No Data Source:} We relied solely on the LLM's internal knowledge without providing additional data.
    \item \emph{3GPP Standards Documents:} We sampled 29 documents from the 3GPP standards repository\footnote{https://www.3gpp.org/} using snowball sampling. The root document with ID 32.260, titled "Telecommunication management - Charging management - IP Multimedia Subsystem (IMS)", was identified by a domain expert from Ericsson with extensive experience in telecommunication standards. The document specifies the standards that a charging management product should comply with.
    \item \emph{3GPP Standards Document with focus on 32.260:} we provided the LLM a similar context as in the previous strategy, but with 32.260 as the main document to extract the classes of the taxonomy.
\end{enumerate}

Using the three data sources and three prompting strategies, we generated nine candidate taxonomies for the telecommunication charging domain. The total number of nodes per strategy is summarized in Table~\ref{tab:taxonomies}.

We presented all taxonomies to the company’s Chief System Architect, a domain expert with over 15 years of experience in telecommunication charging systems. The expert evaluated the taxonomies based on structural coherence, domain relevance, and completeness. While $T_{1x}$ and $T_{3x}$ showed some relevance to the domain, $T_{1x}$ contained many nodes outside the product scope, and $T_{3x}$ exhibited poor structural organization with missing entities. $T_{2x}$ emerged as the strongest candidate, containing the most relevant nodes with a comparatively better hierarchical structure. Ultimately, the expert selected $T_{2a}$ and $T_{2c}$ as the most promising candidates. While both taxonomies demonstrated strong alignment with the domain,  $T_{2a}$ was flagged for two key issues during review:

\begin{enumerate}
    \item \emph{Incorrect Node Placement}: Certain nodes, though domain-relevant, were assigned to inappropriate branches.
    \item \emph{Missing Nodes}: Critical entities were absent given the context of their parent/sibling nodes.
\end{enumerate}

Guided by these findings, the first author refined $T_{2c}$, addressing analogous issues identified in $T_{2a}$ and removing duplicate nodes. The final taxonomy, $T_{2c}$, comprised 675 nodes and was validated by the expert as sufficiently comprehensive and structurally sound for downstream use in TTL creation.

\begin{tcolorbox}[colback=black!5!, colframe=black!15, 
  title=RQ1 How feasible is it to implement TTL in a domain with no well-known established
taxonomy?, fonttitle=\bfseries, coltitle=black, 
  boxrule=0.6mm, arc=2mm, left=1mm, right=1mm, top=1mm, bottom=1mm]
A taxonomy is an essential part of a taxonomic trace links solution. By having sufficient domain-specific documents and utilizing LLMs, it is possible to generate a taxonomy that captures domain concepts on a level that makes it useful to generate taxonomic trace links. Consequently, in a domain without a pre-defined taxonomy, it is still possible to implement TTL if enough domain documents are available.
\end{tcolorbox}

\subsection{Trace Links Creation \& Evaluation}

\begin{figure}[htb]
\begin{subfigure}{.45\textwidth}
  \centering
  \includegraphics[width=1\linewidth]{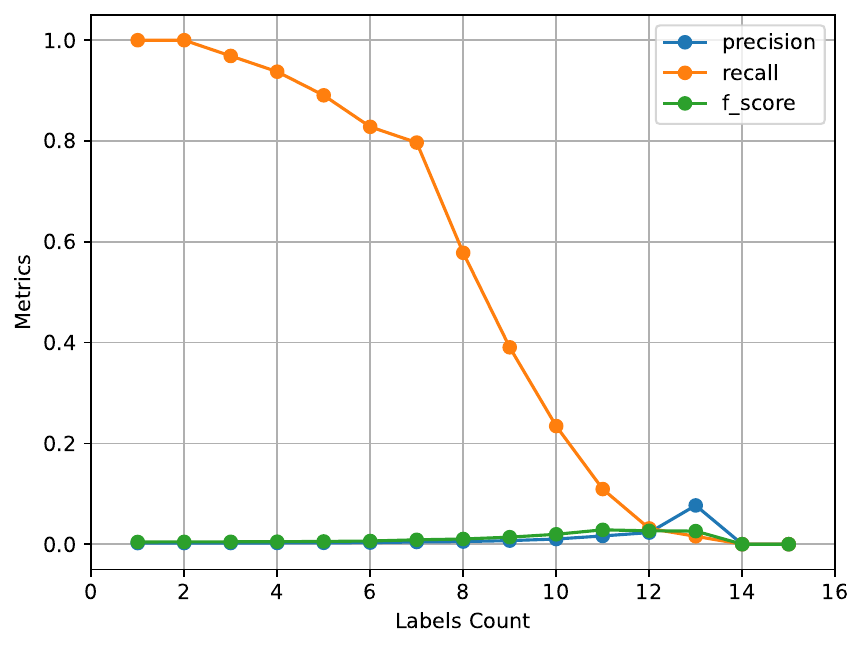}
  \caption{Sentence-T5-XL}
  \label{fig:sentence-t5-xl}
\end{subfigure}
\begin{subfigure}{.45\textwidth}
  \centering
  \includegraphics[width=1\linewidth]{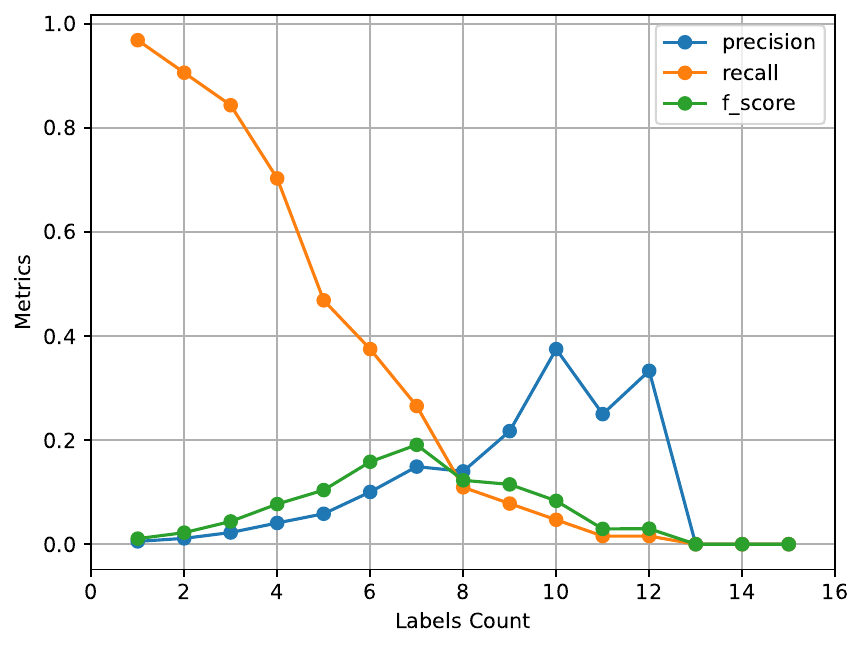}
  \caption{All-MiniLM-L12-v2}
  \label{fig:All-MiniLM-L12-v2}
\end{subfigure}
\caption{Performance of the Zero-Shot classifier to generate TTL}
\label{fig:ttl performance}
\end{figure}

To create trace links, we used a zero-shot classifier~\citep{abdeen2025language} to classify source and target artifacts---against the telecommunication charging taxonomy that we built---and consequently create trace links between them. We compared two state-of-the-art language models: \emph{Sentence-T5-XL} and \emph{All-MiniLM-L12-v2}, which achieved the best performance in the previous study. Trace link candidates were created by matching predicted labels (classes from the taxonomy) between each source-target pair. A trace link candidate is established between a pair if they share \emph{n} matching labels, where \emph{n} (label count, LC) is varied from 1 to 15.
Figure~\ref{fig:ttl performance} summarizes the results for precision, recall, and F1-score.

For \emph{Sentence-T5-XL} (Figure~\ref{fig:sentence-t5-xl}), recall peaked at 1.0 when $LC=1$ \& $LC=2$ but declined as LC increased, reaching 0 at $LC=14$. Precision remained near zero except for LC values between 8–13, where it ranged from 1\% to 8\%. The F1-score was consistently low (0–3\%), indicating no utility. Notably, 97\% of all possible trace links were recommended at $LC=1$, rendering the results ineffective---users would need to manually inspect nearly all target artifacts to identify valid links. Moreover, F1-Score was also low and ranged between 0\%-3\%.

In contrast, \emph{All-MiniLM-L12-v2} (Figure~\ref{fig:All-MiniLM-L12-v2}) achieved marginally lower recall (97\% at $LC=1$, declining to 0 at $LC=13$) but significantly higher precision (1\%–38\%) and F1-scores (1\%–19\%) for LC values between 1–12. While precision remained low, this model demonstrated a better trade-off between recall and precision compared to \emph{Sentence-T5-XL}.

\begin{table}[bth]
    \centering
    \footnotesize
    \caption{Trace Links Candidates Statistics per Artifact Element (LC = 2)}
    \begin{tabular}{p{0.16\textwidth}ccp{0.16\textwidth}p{0.16\textwidth}p{0.16\textwidth}}    
    \toprule
         Scenario & Source & Target & Recommended Links Mean & Recommended Links Standard Deviation & Possible Links per BUC \\
    \midrule
         Scenario 1 & BUC & GPR  & 47 & 33.73 &  277 \\
         Scenario 2 & BUC & BUC  & 114 & 49.67 &  462 \\
    \bottomrule 
    \end{tabular}    
    \label{tab:links_stats}
\end{table}

Based on these evaluation results, we chose \emph{All-MiniLM-L12-v2} and fixed LC at 2, due to the following reasons. First, our intention in using a classifier is to motivate practitioners to create trace links, which we do by recommending a set of links that they need to vet. If the classifier misses many true links (low recall), practitioners will lose trust in the recommender, and they will need to look at all link candidates. On the other hand, if the classifier recommended all possible links (zero precision), the classifier would not provide any reduction in effort. At a high recall, \emph{Sentence-T5-XL} recommended almost all possible links while \emph{All-MiniLM-L12-v2} had relatively better precision. Furthermore, at $LC=2$ (Figure~\ref{fig:All-MiniLM-L12-v2}) recall was at 91\% and precision about 1\%. Although the precision was low, the classifier recommended $\approx$17\% of all possible trace links, a decrease by 83\%. The statistics about the created trace links for each scenario are presented in Table~\ref{tab:links_stats}. The \emph{trace link candidates} refer to the links recommended to the practitioner by matching the labels of the source and target artifacts. While the \emph{possible links} refers to all possible links between the source and target artifacts---without using the classifier---that a practitioner should vet to create correct trace links to perform their task.

Even though the recall of the classifier at $LC=2$ was 91\%, precision was very low at 1\% (Figure~\ref{fig:All-MiniLM-L12-v2}). This results in a very high number of false positives to be vetted by practitioners. Generally, such a classifier is considered unusable. However, as discussed by Berry~\citep{berry_empirical_2021}, the classification task can be considered a ``hairy'' RE task, which is manageable by humans on a small scale but unmanageable on a large scale. In this case, using traditional metrics --- precision, recall, and $F_{1}$-score --- to judge the fit of the classifier for the purpose is inadequate. Instead, researchers need to determine which aspect is most important --- precision or recall --- for the RE task and the tool (classifier)~\citep{berry_panel_2017}. In our use case, we prioritize recall, as it is more important to obtain a complete set of links, including false positives (which are vetted by humans), rather than obtaining fewer false positives but missing many true links (false negatives).

\begin{tcolorbox}[colback=black!5!, colframe=black!15, 
  title={RQ2 What is the performance of the ZSL classifier in recovering trace links between use cases and test cases, measured in terms of precision, recall, and F1-score?}, fonttitle=\bfseries, coltitle=black, 
  boxrule=0.6mm, arc=2mm, left=1mm, right=1mm, top=1mm, bottom=1mm]
The performance results of trace links recovery using the ZSL classifier are presented in Figure~\ref{fig:ttl performance}. \emph{All-MiniLM-L12-v2} achieved better results than \emph{Sentence-T5-XL}. The performance varies depending on the number of matching labels assigned to the traced artifact.
\end{tcolorbox}

\subsection{Traceability Scenarios Evaluation}\label{sec:eval_results}

\begin{figure}
    \centering
    \includegraphics[width=1\linewidth]{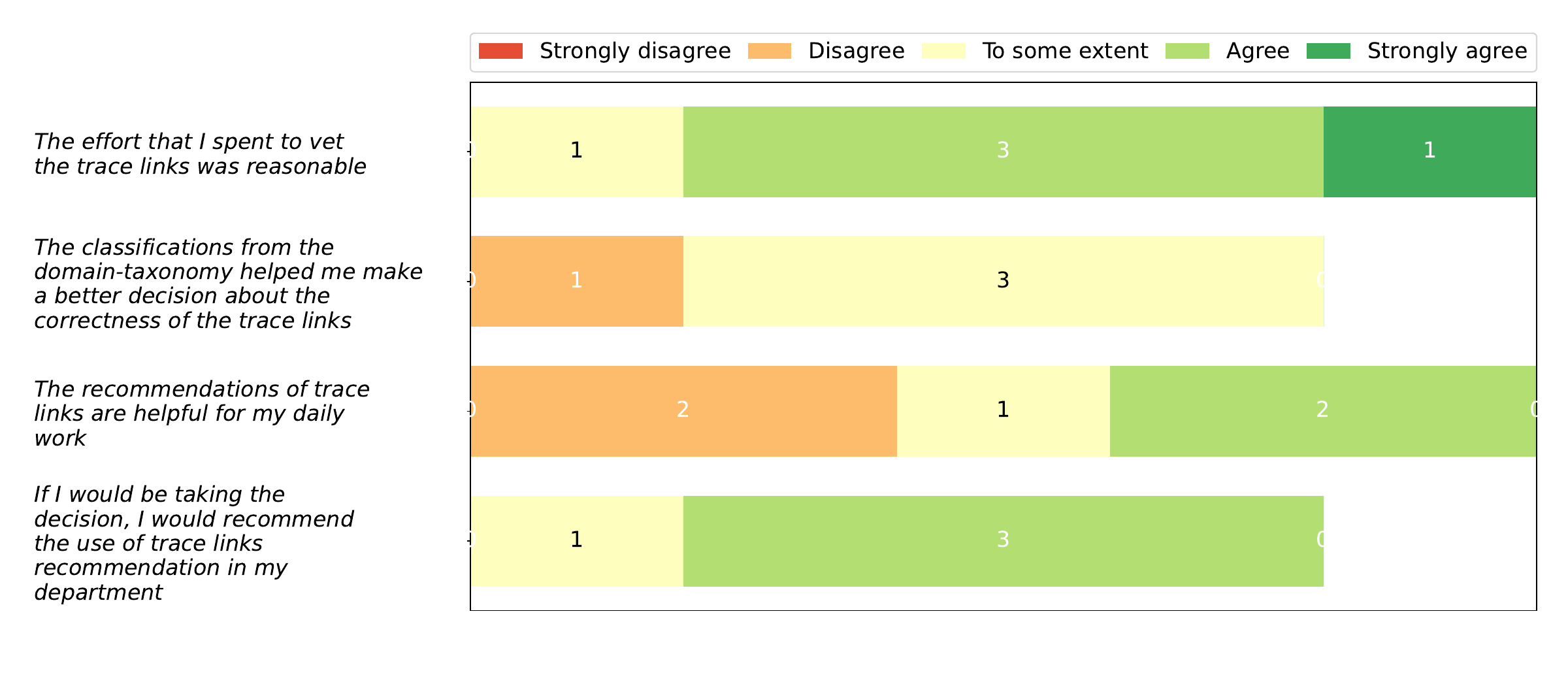}
    \caption{Questionnaire Quantitative Results}
    \label{fig:questionnare}
\end{figure}

To evaluate the strengths and weaknesses of TTL, we conducted two focus group sessions assessing scenarios where traceability is a prerequisite. Participants vetted the trace link candidates for the purpose of compliance checks and dependency identification (see the scenarios identified in Section~\ref{sec:identified-scenarios}). They subsequently answered a mixed-format questionnaire (Likert-scale and open-ended questions) to gauge perceived effort, the usefulness of trace links, and adoption potential. Quantitative responses were analyzed to identify trends, while qualitative feedback provided contextual insights into participants’ reasoning. Figure~\ref{fig:questionnare} summarizes these results, with detailed thematic findings discussed next.

\subsubsection{Human Effort to Validate Trace Link Candidates}
Four out of five participants agreed or strongly agreed that the effort required to vet trace link candidates (up to 30 minutes per trace link set) was reasonable, while one remained neutral. During the focus group session, each participant analyzed 2-3 BUCs. The effort metrics varied by scenario. For \emph{Software Compliance}, participants reviewed 47 trace link candidates (ranging from 7 to 75 out of 277 possible links) between Business Use Cases (BUCs) and domain standards (GPRs). For \emph{Dependencies Identification}, participants evaluated 114 trace link candidates (ranging between 70 and 212 of 462 possible links) between BUCs. To reduce cognitive load, participants prioritized vetting links within \emph{predefined architectural clusters}—logical groupings of related BUCs established by system architects. In daily work, the effort spent to vet trace links of one artifact (30 minutes) could be seen as a burden, especially if the participant creates artifacts more frequently.

\subsubsection{Trace Links Classifications}
Participants received trace links labeled with matched taxonomy classes (trace link candidates) between source and target artifacts. When asked if these classifications aided trace link validation, most were neutral (3/5), while one disagreed. Notably, all participants were unfamiliar with TTL prior to the session. Despite a 30-minute training segment, some reported needing additional time to contextualize the taxonomy’s structure and origin. For instance, one participant remarked, ``I did not have the same pre-understanding as the others'' (P3), highlighting the learning curve associated with domain-specific taxonomies.

\subsubsection{Usefulness of Trace Links Recommendations}\label{sec:usefulness}
Participants expressed mixed perceptions of TTL’s usefulness in daily work: two found the trace links helpful, two found them unhelpful, and one remained neutral.

The participants who did not find the links useful were from both scenarios. In one scenario, \emph{Software Compliance}, the participants stated that they didn't find a value in connecting functional BUCs to GPRs, since, in their opinion, the GPRs are generally only relevant to non-functional BUCs. The participants in the \emph{Software Compliance} scenario were not part of the first focus group session (where the scenarios were identified). Instead, the requirements team was involved in the first session and wanted to have a compliance report per BUC (functional or non-functional). That could explain the observations from the compliance team that the trace links were not useful. Otherwise, we attribute this to BUCs sampling strategy, where most of the selected BUCs describe functional requirements and have only an implicit connection to quality requirements. As stated in Section~\ref{sec:identified-scenarios}, GPRs describe non-functional quality requirements.
In the other scenario, one participant perceived the links as not useful mainly due to the \emph{Multiple Requirements Structures}. Over the past few years, the structure and way of writing requirements have undergone numerous improvements. Consequently, tracing between BUCs with various structures can be difficult.

The participants who found the trace link candidates useful were also from both scenarios. They mentioned three benefits of the recommended links: \emph{Explicit Trace Links}, \emph{Early Trace Links}, and \emph{Undiscovered Trace Links}. In the \emph{Software Compliance} scenario, one participant found it useful to have these links in order to create~\emph{Explicit Trace Links}, mainly due to introducing documentation of explicit trace links between BUCs and GPRs during software compliance analysis in the department, where trace link recommendations can be helpful in reducing the analysis effort. In the \emph{Dependencies Identification} scenario, the participants found the solution useful for \emph{Early Trace Link} creation, i.e., when writing a requirement, as one participant said: \emph{``when creating a new requirement I can look at this to see if they are already covered or dependencies I need to consider''}. Furthermore, the participants were able to find \emph{Undiscovered Trace Links}, as one of the participants mentioned \emph{``It gives a first assessment of possible relationships that might not have been identified already''}. When further asked to elaborate, the participants mentioned that some requirements are old and they may not be able to recall them when writing a new one.

\subsubsection{Decision to Adopt TTL}

The participants were mainly positive (3) or neutral (1) when it came to recommending the solution to be implemented in their department. One participant mentioned that the solution could help \emph{Simplifying Daily Work} mainly by identifying requirements that have correlation, contradiction, or non-compliance. 

\begin{tcolorbox}[colback=black!5!, colframe=black!15, 
  title={RQ3 What is the practical utility of TTL for supporting different traceability use cases?}, fonttitle=\bfseries, coltitle=black, 
  boxrule=0.6mm, arc=2mm, left=1mm, right=1mm, top=1mm, bottom=1mm]
  TTL fell short when identifying relationships between GDRPs and BUCs for \emph{Software Compliance} check. However, the recommended TTLs were useful for \emph{Dependencies Identification} between BUCs. However, implementing TTL in practice requires improving the performance of the zero-shot classifier to make the solution practical.
\end{tcolorbox}

\subsection{Comparison with Existing Practices at the Company} 
Without any trace links recommendations, both scenarios were performed inconsistently at the company, depending on the experience and knowledge of the practitioners in the product. In \emph{Dependency Identification}, practitioners would typically identify dependency based on their experience when they are done writing a BUC. Although this is feasible and common practice, it doesn't guarantee the identification of all dependencies, especially those related written by previous employees. As for \emph{Software Compliance}, the links are currently not explicitly created, and compliance checks are performed only on high-risk BUCs. However, a new policy at the company requires that all BUCs be checked against GPRs, and a compliance check to be carried out. TTL would enable the reduction of manual effort in this activity. However, further studies are required to understand whether the recall in this scenario is sufficiently high to rely on the trace link candidates.  

TTL would aid practitioners in performing these scenarios more systematically by ensuring that a set of potentially relevant trace links is inspected. In \emph{Dependency Identification}, TTL reduces the reliance on the practitioner's experience with the product and ensures that dependencies on old BUCs that may get forgotten are considered. In \emph{Software Compliance}, TTL encourages the practitioner to document the trace links and provide a report of the checked GPRs.

\section{Discussion}\label{sec:discussion}
We discuss the strengths and weaknesses of TTL in industrial traceability scenarios.

\subsection{Observed Limitations}

\paragraph{Multiple Structures of Artifacts} is a barrier to linking artifacts.
TTL aims to unify traceability across artifacts with varying structures by imposing a shared taxonomic framework. While effective for newly created artifacts, linking legacy artifacts remains challenging due to the evolution of terminology and structural changes. For example, participants in the \emph{Dependencies Identification} scenario struggled to validate links between BUCs authored under different structural conventions (Section~\ref{sec:eval_results}). Although the practitioners found it sufficient to document dependencies at the BUC level, structural variability made it difficult to create those links. Establishing links at a finer granularity (e.g., paragraph or sentence) would improve the comparability of artifacts and simplify the creation of trace links. Using traditional trace links and manually identifying the dependency relationship based on experience would, however, not lead to better results; the challenge of linking artifacts with multiple structures will still exist. TTL could reduce the structural mismatch by helping practitioners create new artifacts using a unified terminology and structure, referencing the taxonomy as a guide.

\paragraph{Building a Domain-Specific Taxonomy} requires expertise and good knowledge about the domain. The taxonomy that is required to create trace links should represent the problem domain. Regardless of whether we choose to use automated approaches (e.g., LLMs) or create the taxonomy manually, experts in the domain are necessary to identify a relevant data corpus and the scope of the taxonomy, and ensure the taxonomy is complete. Furthermore, to the best of our knowledge, there is no well-known automated approach to capturing domain knowledge in a structured form. We have experimented with two automated approaches before attempting to use LLMs, namely TaxoGen~\citep{zhang_taxogen_2018} and TaxoCom~\citep{lee_taxocom_2022}. Both approaches resulted in a small taxonomy with a few classes that are mostly irrelevant to the domain. Although using LLMs resulted in more relevant taxonomies, they required post-processing to remove duplicates and fill in the gaps where nodes are missing. Furthermore, we observed that participants in the scenarios, who were unfamiliar with the taxonomy, did not understand why certain labels from the taxonomy were selected to determine the candidate trace links (Section~\ref{sec:usefulness}). The recommender did not provide a \emph{rationale} for the classification. Hence, TTL might be less effective in scenarios where a taxonomy needs to be created first, as opposed to scenarios where taxonomies already exist and are well-known to engineers. 

\paragraph{Dependency on Automation}
TTL is highly dependent on machine-supported classification to create and maintain trace links, mainly due to the large number of taxonomy classes. As presented in Section~\ref{sec:taxonomy-results}, the resulting taxonomy contains 675 classes, and manually classifying artifacts using the taxonomy would be effort-intensive. Thus, TTL is not feasible in practice without some level of automation. As presented in Section~\ref{sec:eval_results}, the effort of trace link creation is significantly reduced due to using the classifier; however, the classifier's performance still needs to be improved to make it useful in practice and allow for the adoption of TTL.

\subsection{Observed Benefits}

\paragraph{Systematic Implementation} of specific traceability scenarios is one of the main opportunities that TTL brings, as presented in Section~\ref{sec:eval_results}. We demonstrated that it is possible to implement TTL in a setting without pre-existing taxonomy, with the help of LLMs and given sufficient domain data (Section~\ref{sec:taxonomy-results}). Once established, TTL leverages pre-trained language models (e.g., \emph{All-MiniLM-L12-v2}) for zero-shot classification. As for traceability maintenance, a change in an artifact requires a review of the labels assigned to the modified artifact only. Then, the trace links can be automatically deduced based on label matching.

\paragraph{Early Trace Links} becomes possible with TTL, as it enables artifact and trace link co-creation. One of the participants noted TTL’s utility in ``checking dependencies when drafting new requirements'' (Section~\ref{sec:eval_results}), underscoring its alignment with developer motivation. Early linking increases trace link completeness, as engineers are incentivized to validate recommendations while contextual knowledge is fresh.
This aligns with \citet{unterkalmsteiner_early_2020}’s vision of proactive traceability.  

\paragraph{Undiscovered Trace Links} were found in the recommended taxonomic trace links (Section~\ref{sec:usefulness}). These links primarily connected the sampled requirements to older requirements that were previously presented and implemented. When creating traditional trace links, these links may be forgotten unless the engineer reviews all documented requirements, which, at least in our case, was not possible. We had a set of 463 BUCs that span just over two years; checking each requirement to find those that are relevant would be impractical. 

\subsection{Threats to Validity}

We discuss the threats to the validity of this study using the framework proposed by Runeson and Höst~\citep{runeson_guidelines_2008}, which includes construct validity, internal validity, external validity, and reliability.

\subsubsection{Construct Validity}
Construct validity concerns the extent to which the study design aligns with the research questions. To mitigate this threat, we iteratively refined the study protocol with input from all authors. The first author drafted the initial protocol, and subsequent revisions incorporated feedback from all co-authors. Additionally, feedback sessions were conducted between the authors to ensure alignment between the study objectives, research questions, and methodology. These sessions helped validate that the chosen methods (e.g., focus groups, taxonomy generation, trace link evaluation) directly address the research questions.

Another threat to construct validity arises from the design of the questionnaire in the focus group session. All items were formulated positively, and the Likert scale used did not include a fully neutral option, which may have biased responses toward agreement. This design choice risks introducing acquiescence and central tendency bias, thereby potentially inflating favourable evaluations. Moreover, the limited number of items reduces the ability to assess internal consistency.

\subsubsection{Internal Validity}
Internal validity concerns the extent to which researcher bias may have influenced the study’s execution. To address this, we implemented the following safeguards: 1) Focus group moderation: at least two researchers were involved in each focus group session, with one researcher leading the discussion and the other observing and asking follow-up questions. This approach reduced the risk of individual bias affecting the outcomes. 2) Stakeholder prioritization: Stakeholders directly prioritized traceability scenarios based on practical relevance to the company, ensuring that the study focused on real-world needs rather than researcher assumptions.

\subsubsection{External Validity} 
Concerns the extent to which the results of the study can be generalized. This study is a single case study conducted in a specific industrial context, and we do not assume that the results are generalizable to all domains. Instead, the primary purpose of this study is to investigate a phenomenon in its natural setting --- namely, the practical implementation of Taxonomic Trace Links (TTL) in a specific setting. While the conclusions from this implementation are not directly transferable to other settings, they nevertheless inform about the characteristics of the approach. The inclusion of additional companies or domains was not feasible due to the nature of TTL, which relies on a domain-specific taxonomy. The results are inherently influenced by the used taxonomy, and traceability scenarios in other domains may differ significantly. Therefore, while the findings provide valuable insights into the application of TTL in the telecommunications domain, caution should be exercised when generalizing them to other contexts.

\subsubsection{Reliability}
Reliability concerns the reproducibility of the study. To enhance reliability, we documented the study protocol, including focus group designs, taxonomy generation steps, and trace link evaluation criteria, in the appendix of this paper. Moreover, all data collection and analysis procedures (e.g., pre-session questionnaires, classification metrics) were systematically recorded to enable replication. The artifacts used to create trace links are proprietary data and are not publicly available, which may affect the replication by external parties. However, replication in other companies requires the use of the company's own data.

\section{Conclusion and Future Work}\label{sec:conclusion}
We evaluated the usefulness of taxonomic trace links (TTL) in an industrial context at Ericsson, using one of its software products, and provided a proof-of-concept for implementing TTL in practice, where a taxonomy does not exist. We illustrated that it is possible to create a taxonomy using LLMs and use it to create TTL. The results of generating a taxonomy using LLMs are promising, as they demonstrate the possibility of adapting TTL in a context without a pre-existing taxonomy to support development activities that are typically time-intensive and rely on various types of trace links. The results of the evaluation of traceability scenarios signal a need to improve the automated generation of trace links using the classifier and trace generator. We identified both the opportunities TTL offers and the limitations that may hinder its adoption in the industry.

In future work, we aim to enhance the precision of the recommender system and address the limitations of implementing TTL in practice---particularly by developing a systematic approach for capturing domain knowledge in a structured taxonomy. Moreover, an investigation into the effort associated with creating and maintaining TTL in comparison to direct trace links should be conducted. TTL links are of a different nature than direct trace links. The links between an artifact (with a size of one page) and taxonomy classes (each with a title of one or a couple of words) are not the same as the links between the same artifact and another one (also with one page size). In other words, vetting whether an engineering artifact is related to a domain concept is presumably easier than vetting whether two engineering artifacts are related to each other. The reasoning behind this is that one needs to have a deep understanding of the content of both artifacts, while the domain concept is presumably part of the vetter's domain knowledge. Thus, the effort can't be calculated by looking \emph{exclusively} at the quantity of links but needs to consider also the nature of the links.
Furthermore, we aim to investigate another identified scenario: regression testing. In this traceability scenario, a set of test cases relevant to an implemented requirement is identified to ensure the changes to the code do not introduce bugs in the system. This scenario is relevant to the testing unit and will be investigated in detail in future studies.

\section{Declaration}

\paragraph{Funding:} This work was funded by the KKS foundation through the SERT Research Profile project (research profile grant 2018/010) at the Blekinge Institute of Technology.

\paragraph{Ethical approval:} Not applicable.

\paragraph{Informed consent:} Not applicable

\paragraph{Author Contributions:} 
\textbf{Waleed Abdeen:} conceptualization, methodology, formal analysis, investigation, and writing - original draft. \textbf{Michael Unterkalmsteiner:} conceptualization, methodology, supervision, writing - review \& editing. \textbf{Peter Löwenadler:} conceptualization, methodology, formal analysis and investigation, writing - review \& editing. \textbf{Parisa Yousefi:} conceptualization, formal analysis, and writing - review \& editing. \textbf{Krzysztof Wnuk:} conceptualization, supervision, and writing - review \& editing. 

\paragraph{Data Availability Statement:} The 3GPP documents used to generate the taxonomy are publicly available here~\footnote{https://www.3gpp.org/specifications-technologies/specifications-by-series}. The traced artifacts are proprietary company data and, therefore, cannot be shared. The domain-specific taxonomy generated in this study will be made available upon a reasonable request and approval from the company. 

\paragraph{Conflict of Interest:} The authors declare that they have no known competing financial interests or personal relationships that could have appeared to influence the work reported in this paper.

\paragraph{Clinical trial number:} Not applicable.

\bibliographystyle{spbasic}      
\bibliography{references}   

\appendix

\section{Scenarios Evaluation Questionnaire}
\label{sec:appendinx-eval}

\setcounter{table}{0}
\renewcommand{\thetable}{\Alph{section}\arabic{table}}

This appendix contains the questionnaire used during the scenario evaluation focus group sessions. Table~\ref{tab:scenario1-questions} contains the questions asked during \emph{Software Compliance} scenario session, while Table~\ref{tab:scenario2-questions} contains the questions asked during \emph{tab:Dependencies Identification} scenario session.

\begin{table}[h]
    \centering
    \caption{Scenario 1 (Software Compliance) Questions}
    \begin{tabular}{cp{0.7\linewidth}c}
    \toprule
        Id & Question & Type/Answer
        \\
    \midrule
        1
         & The effort that I spent to vet the trace links between the BUCs and GPRS was reasonable 
         & Likert scale
         \\
         2 
         & The classifications from the domain-taxonomy helped me make a better decision about the correctness of the trace links 
         & Likert scale
         \\
         3 
         & The recommendations of trace links between BUCs and GPRs are helpful for my daily work 
         & Likert scale 
         \\
         4 
         & Please explain how could these trace links be useful for you daily work, if you think they are not useful, then explain why not 
         & Open ended
         \\
         5
         & If I would be taking the decision, I would recommend the use of trace links recommendation in my department
         & Likert scale
         \\
         6
         & Comments. Please write any additional comments that you have about the solution or the workshop
         & Open ended
         \\
         \bottomrule
    \end{tabular}
    
    \label{tab:scenario1-questions}
\end{table}

\begin{table}[htb]
    \centering
        \caption{Scenario 2 (Dependencies Identification) Questions}
    \begin{tabular}{cp{0.7\linewidth}c}
    \toprule
        Id & Question & Type/Answer
        \\
    \midrule
         1 
         & The effort that I spent to vet the trace links in between the BUCs was reasonable 
         & Likert scale
         \\
         2 
         & The classifications from the domain-taxonomy helped me make a better decision about the correctness of the trace links
         & Likert scale
         \\
         3 
         & The recommendations of trace links in between BUCs are helpful for my daily work
         & Likert scale 
         \\
         4 
         & Please explain how could these trace links be useful for you daily work, if you think they are not useful, then explain why not.
         & Open ended
         \\
         5
         & If I would be taking the decision template, I would recommend the use of trace links recommendation in my department
         & Likert scale
         \\
         6
         & Comments. Please write any additional comments that you have about the solution or the workshop.
         & Open ended
         \\
    \bottomrule
    \end{tabular}
    \label{tab:scenario2-questions}
\end{table}

\section{3GPP Documents}\label{sec:appendix-3gpp}

Table~\ref{tab:3gpp} lists the 3GPP documents that were used as an input to the LLM to generate the domain specific taxonomy.

\begin{table}[]
    \centering
    \caption{3GGP Documents used in Taxonomy Generation}
    \begin{tabular}{cc}
\toprule
Id & How was the document identified? \\
\midrule
32240-i60 & sampled by a domain expert \\
32250-i00 & referenced in 32240 \\
32251-i00 & referenced in 32240 \\
32253-i00 & referenced in 32240 \\
32254-i30& referenced in 32240 \\
32255-j00& referenced in 32240 \\
32256-j00& referenced in 32240 \\
32257-i10& referenced in 32240 \\
32260-i30& referenced in 32240 \\
32270-i30& referenced in 32240 \\
32271-i00& referenced in 32240 \\
32272-i00& referenced in 32240 \\
32273-i00& referenced in 32240 \\
32274-i00& referenced in 32240 \\
32275-i00& referenced in 32240 \\
32276-i00& referenced in 32240 \\
32277-i10& referenced in 32240 \\
32278-i00& referenced in 32240 \\
32280-i00& referenced in 32240 \\
32281-i00& referenced in 32240 \\
32282-i10& referenced in 32240 \\
32290-i50& referenced in 32240 \\
32291-i50& referenced in 32240 \\
32293-i00& referenced in 32240 \\
32295-i00& referenced in 32240 \\
32296-i00& referenced in 32240 \\
32297-i00& referenced in 32240 \\
32298-gd0& referenced in 32240 \\
32299-h10& referenced in 32240 \\
\bottomrule
    \end{tabular}
    \label{tab:3gpp}
\end{table}

\end{document}